\documentclass[11pt,a4paper,fleqn]{article}

\usepackage{amsmath,amssymb,amsthm,enumerate}

\newcounter{mcasenum}
\renewcommand{\themcasenum}{{\rm\arabic{mcasenum}}}

\newtheorem{theorem}{Theorem}
\newtheorem{lemma}{Lemma}

{
\theoremstyle{definition}
\newtheorem{definition}{Definition}

\newtheorem{note}{Note}
\newtheorem*{note*}{Note}
}

\begin{document}

\par\noindent {\LARGE\bf Conservation Laws of Variable Coefficient\\Diffusion--Convection Equations\par}

{\vspace{5mm}\par\noindent {\it
N.M. Ivanova~$^\dag$, R.O. Popovych~$^\ddag$ and C. Sophocleous~$^\star$} \par\vspace{2mm}\par}

{\vspace{2mm}\par\noindent {\it
$^{\dag\hspace{-0.2mm},\ddag}$Institute of Mathematics of NAS of Ukraine,
3 Tereshchenkivska Str., Kyiv-4, 01601, Ukraine\\
\vspace{1em}
E-mail: $^\dag$rop@imath.kiev.ua, $^\ddag$ivanova@imath.kiev.ua \par}

{\vspace{2mm}\par\noindent {\it
$^\star$\,\,Department of Mathematics and Statistics,
University of Cyprus,
CY 1678 Nicosia, Cyprus} \par}
{\par\noindent {$\phantom{\dag}$~\rm E-mail: }{\it
christod@ucy.ac.cy} \par}

{\vspace{6mm}\par\noindent\hspace*{10mm}\parbox{140mm}{\small
We study local conservation laws of variable coefficient
diffusion--convection equations of the form
$f(x)u_t=(g(x)A(u)u_x)_x+h(x)B(u)u_x$. The main tool of our
investigation is the notion of equivalence of conservation laws
with respect to the equivalence groups. That is why, for the class
under consideration  we first construct the usual equivalence
group~$G^{\sim}$ and the extended one~$\hat G^{\sim}$ including
transformations which are nonlocal with respect to arbitrary
elements. The~extended equivalence group~$\hat G^{\sim}$ has
interesting structure since it contains a non-trivial subgroup of
gauge equivalence transformations. Then, using the most direct
method, we carry out two classifications of local conservation
laws up to equivalence relations generated by~$G^{\sim}$ and $\hat
G^{\sim}$, respectively. Equivalence with respect to $\hat
G^{\sim}$ plays the major role for simple and clear formulation of
the final results.
%
}

\section{Introduction}

In this paper we study local conservation laws of PDEs of the general form
\begin{equation} \label{eqDKfgh}
f(x)u_t=(g(x)A(u)u_x)_x+h(x)B(u)u_x,
\end{equation}
where $f=f(x),$ $g=g(x),$ $h=h(x),$ $A=A(u)$ and $B=B(u)$ are arbitrary smooth functions of their variables, and
$f(x)g(x)A(u)\!\neq\! 0.$

Conservation laws were investigated for some subclasses of class~\eqref{eqDKfgh}.
In~particular, Dorodnitsyn and Svirshchevskii~\cite{Dorodnitsyn&Svirshchevskii1983}
(see also~\cite[Chapter~10]{Ibragimov1994V1})
constructed the local conservation laws for the class of reaction--diffusion equations
of the form~$u_t=(A(u)u_x)_x+C(u)$, which has non-empty intersection with the class under consideration.
The first-order local conservation laws of equations~\eqref{eqDKfgh} with $f=g=h=1$
were constructed by Kara and Mahomed~\cite{Kara&Mahomed2002}.
Developing the results obtained in~\cite{Bluman&Doran-Wu1995} for the case~$hB=0$, $f=1$,
in the recent papers~\cite{Ivanova2004,Popovych&Ivanova2004ConsLawsLanl}
we completely classified potential conservation laws (including arbitrary order local ones)
of equations~\eqref{eqDKfgh} with $f=g=h=1$ with respect to the corresponding equivalence group.

For class~\eqref{eqDKfgh} in Section~\ref{SecEqTr} we first construct
the usual equivalence group~$G^{\sim}$ and the extended one~$\hat G^{\sim}$
including transformations which are nonlocal with respect to arbitrary elements.
We discuss the structure of the extended equivalence group~$\hat G^{\sim}$ having
 non-trivial subgroup of gauge equivalence transformations.
Then we carry out two classifications of local conservation laws
up to the equivalence relations generated by~$G^{\sim}$ and $\hat
G^{\sim}$, respectively, using the most direct method
(Section~\ref{SecLCL}).

The main tool of our investigation is the notion of
equivalence of conservation laws with respect to equivalence groups,
which was introduced in~\cite{Popovych&Ivanova2004ConsLawsLanl}.
Below we adduce some necessary notions and statements, restricting ourselves to the case of two independent variables.
See~\cite{Olver1986,Popovych&Ivanova2004ConsLawsLanl} for more details and general formulations.

Let~$\mathcal{L}$ be a system~$L(t,x,u_{(\rho)})=0$ of PDEs
for unknown functions $u=(u^1,\ldots,u^m)$
of independent variables~$t$ (the time variable) and~$x$ (the space variable).
Here $u_{(\rho)}$ denotes the set of all the partial derivatives of the functions $u$
of order no greater than~$\rho$, including $u$ as the derivatives of the zero order.

\begin{definition}\label{def.conservation.law}
A {\em conservation law} of the system~$\mathcal{L}$ is a divergence expression
\begin{equation}\label{conslaw}
D_tF(t,x,u_{(r)})+D_xG(t,x,u_{(r)})=0
\end{equation}
which vanishes for all solutions of~$\mathcal{L}$.
Here $D_t$ and $D_x$ are the operators of total differentiation with respect to $t$ and $x$, respectively;
$F$ and $G$ are correspondingly called the {\em density} and the {\em flux} of the conservation law.
\end{definition}

Two conserved vectors $(F,G)$ and $(F',G')$ are {\em equivalent} if
there exist functions~$\hat F$, $\hat G$ and~$H$ of~$t$, $x$ and derivatives of~$u$ such that
$\hat F$ and $\hat G$ vanish for all solutions of~$\mathcal{L}$~and
$F'=F+\hat F+D_xH$, $G'=G+\hat G-D_tH$.

\begin{lemma}\label{PropositionOnInducedMapping}\cite{Popovych&Ivanova2004ConsLawsLanl}
Any point transformation $g$ between systems~$\mathcal{L}$ and~$\tilde{\mathcal{L}}$
induces a~linear one-to-one mapping $g_*$ between the corresponding linear spaces of conservation laws.
\end{lemma}

Consider the class~$\mathcal{L}|_S$ of systems~$L(t,x,u_{(\rho)},\theta(t,x,u_{(\rho)}))=0$
parameterized with the pa\-ra\-me\-ter-functions~$\theta=\theta(t,x,u_{(\rho)}).$
Here $L$ is a tuple of fixed functions of $t$, $x$, $u_{(\rho)}$ and $\theta$.
$\theta$ denotes the tuple of arbitrary (parametric) functions
$\theta(t,x,u_{(\rho)})=(\theta^1(t,x,u_{(\rho)}),\ldots,\theta^k(t,x,u_{(\rho)}))$
satisfying the additional condition~$S(t,x,u_{(\rho)},\theta_{(q)}(t,x,u_{(\rho)}))=0$.

Let~$P=P(L,S)$ denote the set of pairs each from which consists of
a system from~$\mathcal{L}|_S$ and a conservation law of this system.
Action of transformations from an equivalence group~$G^{\sim}$ of the class~$\mathcal{L}|_S$
together with the pure equivalence relation of conserved vectors
naturally generates an equivalence relation on~$P$.
Classification of conservation laws with respect to~$G^{\sim}$ will be understood as
classification in~$P$ with respect to the above equivalence relation.
This problem can be investigated in the way that it is similar to group classification in classes
of systems of differential equations. Specifically, we firstly construct the conservation laws
that are defined for all values of the arbitrary elements.
(The corresponding conserved vectors may depend on the arbitrary elements.)
Then we classify, with respect to the equivalence group, arbitrary elements for each of the systems
that admits additional conservation laws.

\section{Equivalence transformations and choice of investigated class}\label{SecEqTr}

In order to classify the conservation laws of equations of the class~\eqref{eqDKfgh},
firstly we have to investigate equivalence transformations of this class.

The usual equivalence group~$G^{\sim}$ of class~\eqref{eqDKfgh} is formed by the nondegenerate point transformations
in the space of~$(t,x,u,f,g,h,A,B)$, which are projectible on the space of~$(t,x,u)$,
i.e. they have the form
\begin{gather}
(\tilde t,\tilde x,\tilde u)=(T^t,T^x,T^u)(t,x,u), \nonumber\\[0.5ex]
(\tilde f,\tilde g,\tilde h,\tilde A,\tilde B)=(T^f,T^g,T^h,T^A,T^B)(t,x,u,f,g,h,A,B),\label{EquivTransformations}
\end{gather}
and transform any equation from the class~\eqref{eqDKfgh} for the function $u=u(t,x)$
with the arbitrary elements $(f,g,h,A,B)$
to an equation from the same class for function $\tilde u=\tilde u(\tilde t,\tilde x)$
with the new arbitrary elements~$(\tilde f,\tilde g,\tilde h,\tilde A,\tilde B)$.

\begin{theorem}
$G^{\sim}$ consists of the transformations
\[
\begin{array}{l}
\tilde t=\delta_1 t+\delta_2,\quad
\tilde x=X(x), \quad
\tilde u=\delta_3 u+\delta_4, \\[1ex]
\tilde f=\dfrac{\varepsilon_1\delta_1 f}{X_x(x)}, \quad
\tilde g=\varepsilon_1\varepsilon_2^{-1}X_x(x) g, \quad
\tilde h=\varepsilon_1\varepsilon_3^{-1}h, \quad
\tilde A=\varepsilon_2A, \quad
\tilde B=\varepsilon_3B,
\end{array}
\]
where $\delta_j$ $(j=\overline{1,4})$ and $\varepsilon_i$ $(i=\overline{1,3})$ are arbitrary constants,
$\delta_1\delta_3\varepsilon_1\varepsilon_2\varepsilon_3\not=0$, $X$ is an arbitrary smooth function of~$x$, $X_x\not=0$.
\end{theorem}

It appears that class~\eqref{eqDKfgh} admits other equivalence transformations which do not belong to~$G^{\sim}$
and form, together with usual equivalence transformations, an extended equivalence group.
We demand for these transformations to be point with respect to $(t,x,u)$.
The explicit form of the new arbitrary elements~$(\tilde f,\tilde g,\tilde h,\tilde A,\tilde B)$ is determined
via $(t,x,u,f,g,h,A,B)$ in some non-fixed (possibly, nonlocal) way.
We construct the complete (in this sense) extended equivalence group~$\hat G^{\sim}$ of
class~\eqref{eqDKfgh}, using the direct method.

Existence of such transformations can be explained in many respects by features of representation
of equations in the form~\eqref{eqDKfgh}.
This form leads to an ambiguity since the same equation has an infinite series of
different representations. More exactly, two representations~\eqref{eqDKfgh}
with the arbitrary element tuples $(f,g,h,A,B)$ and $(\tilde f,\tilde g,\tilde h,\tilde A,\tilde B)$
determine the same equation iff
\begin{gather}
\tilde f=\varepsilon_1\varphi f, \quad
\tilde g=\varepsilon_1\varepsilon_2^{-1}\varphi g, \quad
\tilde h=\varepsilon_1\varepsilon_3^{-1}\varphi h, \quad 
\tilde A=\varepsilon_2 A, \quad
\tilde B=\varepsilon_3 (B+\varepsilon_4 A),\label{GaugeEquivTransformationsDKfgh}
\end{gather}
where $\varphi=\exp\left(-\varepsilon_4\int \frac{h(x)}{g(x)}dx\right)$,
$\varepsilon_i$ $(i=\overline{1,4})$ are arbitrary constants, $\varepsilon_1\varepsilon_2\varepsilon_3\not=0$
(the variables $t$, $x$ and $u$ do not transform!).

The transformations~\eqref{GaugeEquivTransformationsDKfgh} act only on arbitrary elements
and do not really change equations.
In general, transformations of such type can be considered as trivial~\cite{LisleDissertation}
(``gauge'') equivalence transformations
and form the ``gauge'' (normal) subgroup~$\hat G^{\sim g}$ of the extended equivalence group~$\hat G^{\sim}$.
Application of ``gauge'' equivalence transformations is equivalent to rewriting equations
in another form. In spite of really equivalence transformations, their role in group classification
comes not as a choice of representatives in equivalence classes but as a choice of the form of these representatives.

Let us note that transformations~\eqref{GaugeEquivTransformationsDKfgh} with $\varepsilon_4\not=0$ are
nonlocal with respect to arbitrary elements, otherwise they belong to~$G^{\sim}$
and form the ``gauge'' (normal) subgroup~$G^{\sim g}$ of the equivalence group~$G^{\sim}$.

The factor-group $\hat G^{\sim}/\hat G^{\sim g}$ coincides for class~\eqref{eqDKfgh} with~$G^{\sim}/G^{\sim g}$
and can be assumed to consist of the transformations
\begin{equation} \label{RealEquivTransformationsDKfgh}\arraycolsep=0em
\begin{array}{l}
\tilde t=\delta_1 t+\delta_2,\quad
\tilde x=X(x), \quad
\tilde u=\delta_3 u+\delta_4,\\[1ex]
\tilde f=\dfrac{\delta_1 f}{X_x(x)}, \quad
\tilde g=X_x(x) g, \quad
\tilde h=h, \quad
\tilde A=A, \quad
\tilde B=B,
\end{array}
\end{equation}
where $\delta_i$ ($i=\overline{1,4}$) are arbitrary constants, $\delta_1\delta_3\not=0$,
$X$ is an arbitrary smooth function of~$x$, $X_x\not=0$.

Using the transformation $\tilde t=t$, $\tilde x=\int \frac{dx}{g(x)}$, $\tilde u=u$
from $G^{\sim}/G^{\sim g}$, we can reduce equation~(\ref{eqDKfgh}) to
$
\tilde f(\tilde x)\tilde u_{\tilde t}= (A(\tilde u)
\tilde u_{\tilde x})_{\tilde x} + \tilde h(\tilde x)B(\tilde u)\tilde u_{\tilde x},
$
where $\tilde f(\tilde x)=g(x)f(x)$, $\tilde g(\tilde x)=1$ and $\tilde h(\tilde x)=h(x)$.
(Likewise any equation of form~\eqref{eqDKfgh} can be reduced to the same form with $\tilde f(\tilde x)=1.$)
That is why, without loss of generality we restrict ourselves to investigation of the equation
\begin{equation} \label{eqDKfh}
f(x)u_t=\left(A(u)u_x \right)_x + h(x)B(u)u_x.
\end{equation}

Any transformation from~$\hat G^{\sim}$, which preserves the condition $g = 1$, has the form
\begin{equation} \label{EquivTransformationsDKfh}\arraycolsep=0em
\begin{array}{l}
\tilde t=\delta_1 t+\delta_2,\quad
\tilde x=\delta_5 \int e^{\delta_8\int\! h\,dx}dx+\delta_6, \quad
\tilde u=\delta_3 u+\delta_4,\quad 
\tilde f=\delta_1\delta_5^{-1}\delta_9 fe^{-2\delta_8\int\! h\,dx}, \\[1ex]
\tilde h=\delta_9\delta_7^{-1} he^{-\delta_8\int\! h\,dx}, \quad 
\tilde A=\delta_5\delta_9A, \quad
\tilde B=\delta_7(B+\delta_8A),
\end{array}
\end{equation}
where $\delta_i$ ($i=\overline{1,9}$) are arbitrary constants, $\delta_1\delta_3\delta_5\delta_7\delta_9\not=0$.
The set~$\hat G^{\sim}_1$ of such transformations
is a subgroup of~$\hat G^{\sim}$. It can be considered as a generalized equivalence group of class~\eqref{eqDKfh} after
admitting dependence of~\eqref{EquivTransformations} on arbitrary elements~\cite{Meleshko1994}
and additional supposition that such dependence can be nonlocal.
The group~$G^{\sim}_1$ of usual (local) equivalence transformations of class~\eqref{eqDKfh}
coincides with the subgroup singled out from~$\hat G^{\sim}_1$ via the condition $\delta_8 = 0$.
The transformations~\eqref{EquivTransformationsDKfh} with non-vanishing values of the parameter $\delta_8$
are nonlocal and are compositions of
(nonlocal) gauge and usual equivalence transformations from~$G^{\sim}_1$.

There exists a way to avoid operations with nonlocal in $(t, x, u)$ equivalence transformations.
More exactly, we can assumed that the parameter-function $B$ is determined up to an additive
term proportional to $A$ and subtract such term from $B$ before applying equivalence
transformations~\eqref{RealEquivTransformationsDKfgh}.

\section{Local conservation laws}\label{SecLCL}

We search local conservation laws of equations from class~(\ref{eqDKfh}).

\begin{lemma}\label{LemmaOnOrderOfConsLawsOfDCEs}
Any conservation law of form~\eqref{conslaw} of any equation from
class~\eqref{eqDKfh} is equivalent to a conservation law that has
the density depending on $t$, $x$, and $u$ and the flux depending
on $t$, $x$, $u$ and $u_x$.
\end{lemma}

\begin{note}
A similar statement is true for an arbitrary (1+1)-dimensional evolution equation~$\cal L$ of the even
order~$r=2\bar r$, $\bar r\in\mathbb{N}$.
For example~\cite{Ibragimov1985}, for any conservation law of~$\cal L$
we can assume up to equivalence of conserved vectors
that~$F$ and $G$ depend only on~$t$, $x$ and derivatives of~$u$ with respect to~$x$, and
the maximal order of derivatives in~$F$ is not greater than $\bar r$.
\end{note}

\setcounter{mcasenum}{0}

\begin{theorem}\label{TheorClassSmallGroup}
A complete list of $G^{\sim}_1$-inequivalent equations~\eqref{eqDKfh} having nontrivial
conservation laws is exhausted by the following ones
\begin{gather*}\textstyle
\makebox[7mm][l]{\refstepcounter{mcasenum}\themcasenum\label{h1}.}
 h=1 :\quad (\,fu,\ -Au_x-\int\!\! B\,).\\[0.5ex]\textstyle
\makebox[7mm][l]{\refstepcounter{mcasenum}\themcasenum\label{hx-1}.} h=x^{-1} :\quad
(\,xfu,\ -xAu_x+\int\!\!A-\int\!\!B\,).\\[0.5ex]\textstyle
\makebox[7mm][l]{\refstepcounter{mcasenum}\themcasenum\label{B0}.}
B=\varepsilon A: \quad
(yfe^{-\varepsilon\!\int\!\! h} u,\ -yAe^{-\varepsilon\!\int\!\! h}u_y+\int\!\! A),
\  (fe^{-\varepsilon\!\int\!\! h}u,\ -Ae^{-\varepsilon\!\int\!\! h}u_y).\\[0.5ex]
\makebox[7mm][l]{\refstepcounter{mcasenum}\themcasenum\label{B1fhint}.}
B=\varepsilon A+1, \quad f=-hZ^{-1},
\quad h=Z^{-1/2}\exp\left(-\int\dfrac{a_{00}+a_{11}}{2Z}dy\right):\\[0.5ex] \textstyle
\makebox[7mm][l]{}(\,(\sigma^{k1}y+\sigma^{k0})fe^{-\varepsilon\int\!\! h}u,\;
-(\sigma^{k1}y+\sigma^{k0})(Ae^{-\varepsilon\int\!\! h}u_y+hu)+\sigma^{k1}\int\!\! A\,)\\[0.5ex] \textstyle
\makebox[7mm][l]{\refstepcounter{mcasenum}\themcasenum\label{B1fhx}.}
 B=\varepsilon A+1, \quad f=h_y: \quad
 (\,e^{t-\varepsilon\int\!\! h}h_yu,\  -e^{t}(Ae^{-\varepsilon\int\!\! h}u_y+hu)\,).
\\[0.5ex] \textstyle
\makebox[7mm][l]{\refstepcounter{mcasenum}\themcasenum\label{B1fhx+hx-1}.} B=\varepsilon A+1, \quad f=h_y+hy^{-1}:
\quad (\,e^{t-\varepsilon\int\!\! h}yfu ,\ -e^t(yAe^{-\varepsilon\int\!\! h}u_y+yhu-\int\!\! A)\,).\\[0.5ex]
\makebox[7mm][l]{\refstepcounter{mcasenum}\themcasenum\label{A1Bne0}.}
 A=1, \quad B_u\ne0,\quad f=-h(h^{-1})_{xx}: 
\quad (\,e^t(h^{-1})_{xx} u,\ e^t(h^{-1}u_x-(h^{-1})_xu+\int\!\! B)\,). \\[0.5ex] \textstyle
\makebox[7mm][l]{\refstepcounter{mcasenum}\themcasenum\label{A1B0}.}
 A=1, \quad B=0: \quad (\,\alpha f u, \ -\alpha u_x+\alpha_x u\,).
\end{gather*}
Here
$y$ is implicitly determined by the formula $x=\int\! e^{\varepsilon \int\!h(y)dy}dy$;
$\varepsilon,a_{ij}=\rm const$, $i,j=\overline{0,1}$;
$(\sigma^{k1},\sigma^{k0})=(\sigma^{k1}(t),\sigma^{k0}(t))$, $k=\overline{1,2}$, is a fundamental solution
of the system of ODEs $\sigma^\nu_t=a_{\mu\nu}\sigma^\mu$;
$Z=a_{01}y^2+(a_{00}-a_{11})y-a_{10}$;
$\alpha=\alpha(t,x)$  is an arbitrary solution of the linear equation
$f\alpha_t+\alpha_{xx}=0$.
(Together with constraints on $A$, $B$, $f$ and $h$ we also adduce complete lists of linear independent
conserved vectors.)
\end{theorem}

In Theorem~\ref{TheorClassSmallGroup} we classify conservation laws
with respect to the usual equivalence group $G^{\sim}_1$.
The results that are obtained can be formulated in an implicit form only,
and indeed Case~4 is split into a number of inequivalent cases depending on values of~$a_{ij}$.
At the same time, using the
extended equivalence group~$\hat G^{\sim}_1$,
we can present the result of classification in a closed and simple form with a smaller number
of inequivalent equations having nontrivial conservation laws.

\begin{theorem}\label{TheorClassWideGroup}
A complete list of $\hat G^{\sim}_1$-inequivalent equations~\eqref{eqDKfh} having nontrivial
conservation laws is exhausted by the following ones
\setcounter{mcasenum}{0}
\begin{gather*}\textstyle
\makebox[9mm][l]{\refstepcounter{mcasenum}\themcasenum\label{2.h1}.}
h=1 :\quad (\,fu,\ -Au_x-\int\!\! B\,).\\[0.7ex]\textstyle
\makebox[9mm][l]{\refstepcounter{mcasenum}\themcasenum\label{2.B0}{\rm a}.}
B=0: \quad  (\,fu,\ -Au_x\,),\
 (\,xfu,\ -xAu_x+\int\!\! A\,).\\[0.7ex] \textstyle
\makebox[9mm][l]{\themcasenum{\rm b}.}
B=1, \quad f=1,\quad h=1: \quad (\,u ,\ -Au_x-u\,), \\ \textstyle
\makebox[9mm][l]{} (\,(x+t)u ,\ -(x+t)(Au_x+u)+\int\!\! A\,).\\[0.7ex] \textstyle
\makebox[9mm][l]{\themcasenum{\rm c}.} B=1, \quad f=e^x,\quad h=e^x:\quad (\,e^{x+t}u ,\ -e^t(Au_x+e^xu)\,),\\ \textstyle
\makebox[9mm][l]{} (\,e^{x+t}(x+t)u ,\ -e^t(x+t)(Au_x+e^xu)+e^t\int\!\! A\,).\\[0.7ex] \textstyle
\makebox[9mm][l]{\themcasenum{\rm d}.} B=1, \quad f=x^{\mu-1},\quad h=x^\mu:\quad
 (\,x^{\mu-1}e^{\mu t}u ,\ -e^{\mu t}(Au_x+x^\mu u)\,),\\ \textstyle
\makebox[9mm][l]{} (\,x^\mu e^{(\mu+1)t}u ,\ e^{(\mu+1)t}(-xAu_x-x^{\mu+1} u+\int\!\! A)\,) .\\[0.7ex] \textstyle
\makebox[9mm][l]{\refstepcounter{mcasenum}\themcasenum\label{2.B1fexphexp}.} B=1,
\quad f=e^{\mu/x}x^{-3},\quad h=e^{\mu/x}x^{-1},\quad \mu\in\{0,1\}: \\ \textstyle
\makebox[9mm][l]{}
(\,fe^{-\mu t}xu ,\ -e^{-\mu t}x(Au_x+hu)+e^{-\mu t}\int\!\! A\,),\\ \textstyle
\makebox[9mm][l]{} (\,fe^{-\mu t}(tx-1)u ,\ -e^{-\mu t}(tx-1)(Au_x+hu)+te^{-\mu t}\int\!\! A\,).\\[0.7ex] \textstyle
\makebox[9mm][l]{\refstepcounter{mcasenum}\themcasenum\label{2.B1fx1x1hx1x1}.} B=1,
\ f=|x-1|^{\mu-3/2}|x+1|^{-\mu-3/2},\ h=|x-1|^{\mu-1/2}|x+1|^{-\mu-1/2}:
\\ \textstyle
\makebox[9mm][l]{}
(\,fe^{(2\mu+1)t}(x-1)u ,\ -e^{(2\mu+1)t}(x-1)(Au_x+hu)+e^{(2\mu+1)t}\int\!\! A\,),\\ \textstyle
\makebox[9mm][l]{} (\,fe^{(2\mu-1)t}(x+1)u ,\ -e^{(2\mu-1)t}(x+1)(Au_x+hu)+e^{(2\mu-1)t}\int\!\! A\,).
\\[0.7ex]\textstyle
\makebox[9mm][l]{\refstepcounter{mcasenum}\themcasenum\label{2.B1farctanharctan}.}
B=1, \quad f=e^{\mu\arctan x}(x^2+1)^{-3/2},
\quad h=e^{\mu\arctan x}(x^2+1)^{-1/2}:\\ \textstyle
\makebox[9mm][l]{} (\,fe^{\mu t}(x\cos t+\sin t)u ,\
-e^{\mu t}(x\cos t+\sin t)(Au_x+hu)+e^{\mu t}\cos t\int\!\! A\,),\\ \textstyle
\makebox[9mm][l]{} (\,fe^{\mu t}(x\sin t-\cos t)u,  \ -e^{\mu t}(x\sin t-\cos t)(Au_x+hu)+e^{\mu t}\sin t\int\!\! A\,).
\\[0.7ex] \textstyle
\makebox[9mm][l]{\refstepcounter{mcasenum}\themcasenum\label{2.B1fhx}.}
B=1, \quad f=h_x: \quad (\,e^th_xu,\ -e^t(Au_x+hu)\,) .
\\[0.7ex] \textstyle
\makebox[9mm][l]{\refstepcounter{mcasenum}\themcasenum\label{2.B1fhx+hx-1}.} B=1, \quad f=h_x+hx^{-1}:
\quad (\,e^txfu ,\ -e^t(xAu_x+xhu-\int\!\! A)\,).\\[0.7ex] \textstyle
\makebox[9mm][l]{\refstepcounter{mcasenum}\themcasenum\label{2.A1Bne0}.}
A=1, \quad B_u\ne0,\quad f=-h(h^{-1})_{xx}: 
\quad (\,e^t(h^{-1})_{xx} u,\ e^t(h^{-1}u_x-(h^{-1})_xu+\int\!\! B)\,). \\[0.7ex] \textstyle
\makebox[9mm][l]{\refstepcounter{mcasenum}\themcasenum\label{2.A1B0}.}
A=1, \quad B=0: \quad (\,\alpha f u, \ -\alpha u_x+\alpha_x u\,).
\end{gather*}
Here $\mu={\rm const}$,
$\alpha=\alpha(t,x)$  is an arbitrary solution of the linear equation \mbox{$f\alpha_t+\alpha_{xx}=0$}.
(Together with constraints on $A$, $B$, $f$ and $h$ we also adduce complete lists of linear independent
conserved vectors.)
\end{theorem}

\begin{note}
The cases 2b--2d can be reduced to the case 2a by means of
additional equivalence transformations:\\[0.5ex]
2b$\to$2a:\quad $\tilde t=t$, $\tilde x=x+t$, $\tilde u=u$;\\[0.5ex]
2c$\to$2a:\quad $\tilde t=e^t$, $\tilde x=x+t$, $\tilde u=u$;\\[0.5ex]
2d ($\mu+1\ne0$)$\to$2a: \quad
$\tilde t=(\mu+1)^{-1}(e^{(\mu+1)t}-1)$,
$\tilde x=e^tx$, $\tilde u=u$;\\[0.5ex]
2d ($\mu+1=0$)$\to$2a:\quad
$\tilde t=t$,
$\tilde x=e^tx$, $\tilde u=u$.
\end{note}

\section{Conclusion}

The present paper is the beginning for further studies on this subject.
For the class
under consideration we intend to perform a complete classification
of potential conservation laws and construct an exhaustive list of
locally inequivalent potential systems corresponding to them.
These results  can be developed and generalized in a number of
different directions. So, studying different kinds of symmetries (Lie,
nonclassical, generalized ones) of constructed potential systems,
we may obtain the corresponding kinds of potential symmetries
(usual potential, nonclassical potential, generalized potential).
Analogously, local equivalence transformations between
potential systems constructed for different initial equations
result in nonlocal (potential) equivalence transformations for the
class under consideration (see e.g.~\cite{Popovych&Ivanova2005PETs}). In such way it is possible to find new
nonlocal connections between variable coefficient
diffusion--convection equations. We believe that the same approach
used in this article, can be employed for investigation of wider
classes of differential equations.

\subsection*{Acknowledgements}

The research of NMI was supported by NAS of Ukraine in the form of the grant for young scientists and by
University of Cyprus. NMI and ROP express their gratitude to the
hospitality shown by the University of Cyprus during their visit to
the University. CS also expresses his gratitude to all members of
the Department of Applied Research of Institute of Mathematics of
NAS of Ukraine for their warm hospitality during his visit to Kyiv.

\end{document}